\documentclass[doublecol]{epl2}
\usepackage{amsmath}
\usepackage{amssymb}
\usepackage{graphics}

\title{Analytical approximation for single-impurity Anderson model}
\shorttitle{Analytical approximation for SIAM}

\author{I.S. Krivenko\inst{1}, A.N. Rubtsov\inst{1}, M.I. Katsnelson \inst{2}  \and A.I. Lichtenstein\inst{3}}

\institute{
  \inst{1} Department of Physics, Moscow State University, 119992 Moscow, Russia\\
  \inst{2} Radboud University, Heijendaalseweg 135, 6525AJ, Nijmegen, The Netherlands\\
  \inst{3} Institut f{\"u}r Theoretische Physik, Universit{\"a}t Hamburg, Jungiusstra{\ss}e 9, D-20355 Hamburg, Germany\\
}


\abstract{We have applied the recently developed dual fermion
technique to the spectral properties of single-band Anderson
impurity problem (SIAM). In our approach a series expansion is
constructed in vertices of the corresponding atomic Hamiltonian
problem. This expansion contains a small parameter in two limiting
cases: in the weak coupling case ($U/t \to 0$), due to the smallness
of the irreducible vertices, and near the atomic limit ($U/t \to
\infty$), when bare propagators are small. Reasonable results are
obtained also for the most interesting case of strong correlations
($U \approx t$). The atomic problem of the Anderson impurity model
has a degenerate ground state, so the application of the
perturbation theory is not straightforward. We construct a special
approach dealing with symmetry-broken ground state of the
renormalized atomic problem. Formulae for the first-order dual
diagram correction are obtained analytically in the real-time
domain. Most of the Kondo-physics is reproduced: logarithmic
contributions to the self energy arise, Kondo-like peak at the Fermi
level appears, and the Friedel sum rule is fulfilled. Our approach
describes also renormalization of atomic resonances due to
hybridization with a conduction band. A generalization of the
proposed scheme to a multi-orbital case can be important for the
realistic description of correlated solids.}

\newcommand{\up}{\uparrow}
\newcommand{\down}{\downarrow}
\newcommand{\sgn}{\mathop{\mathrm{sgn}}\nolimits}

\renewcommand{\Re}{\mathop{\mathrm{Re}}}
\renewcommand{\Im}{\mathop{\mathrm{Im}}}

\begin{document}

\maketitle

\section{Introduction}
The problem of realistic description of spectral properties of
correlated impurity in the metallic surrounding is far from solution
despite of considerable progress during the past years. There are quite
successful tools for large Coulomb $U$, particularly diagrammatic pseudoparticle
approach \cite{wolfle1}. Transport properties can be described in this limit
also \cite{glazman,wolfle2}. The case of moderate $U$ is less studied, although
there is an interesting physics behind, e.g. renormalization of the multiplet
structure due to hybridization with conduction band. An exact solution by
using the Bethe-ansatz \cite{wiegmann,andrei} is obtained for thermodynamical
properties only. There are numerically exact continuous-time quantum Monte Carlo
calculations in imaginary time domain\cite{rubtsov_QMC,werner_QMC}.
Practical accuracy of these methods is limited by an ill-posed problem of
analytical continuation to the real-frequency axis. More accurate data can
be obtained within the Numerical Renormalization Group framework \cite{bulla}.
Finally, it is important to mention a simple RPA-like approach which starts
from the local-moment broken symmetry solution \cite{logan} and describe Kondo
properties surprisingly well. Abrikosov-Suhl resonanse has been reproduced
properly, it fulfils Friedel sum rule and its width corresponds to the Kondo
energy scale.

In the present paper we describe an approach aiming analytical description
of spectral properties of correlated impurity with moderate $U$. In this
first paper we apply our method to a single-band Anderson impurity model (SIAM).
Our approach, which starts from the exact mapping of SIAM to auxiliary (dual)
variables \cite{dual_PRB1,dual_PRB2}, resembles key properties of the 
broken-symmetry self-consistent RPA approach \cite{logan}, but allows fully
analytical consideration. Another important property of our method is a
renormalization procedure, which allows to describe a shift of the atomic
resonances due to hybridization with a conduction band.
\section{Anderson impurity problem}
We consider a single-band Anderson impurity problem at zero temperature. The action is
\begin{eqnarray*}\label{act}
S = S_{at} -\iint_{-\infty}^{+\infty} dt\ dt' \bar c_{\sigma t} \Delta(t-t') c_{\sigma t'}\\
S_{at} = \int_{-\infty}^{+\infty} dt \left( 
        i \bar c_{\sigma t} \frac{\partial}{\partial t} c_{ \sigma t} 
           -  U \tilde{n}_{\up t} \tilde{n}_{\down t} + \mu \tilde{n}_{\sigma t}
            \right)
\end{eqnarray*}
Summations over spin index $\sigma=\up, \down$ are supposed; 
$\tilde n_{\sigma t} \equiv \frac{1}{2}(\bar{c}_{\sigma t-0} c_{\sigma t} +\bar{c}_{\sigma t+0} c_{\sigma t})$
corresponds to $n-\frac{1}{2}$ term in the Hamiltonian \cite{kleinert}, so that there is
a particle-hole symmetry of $S_{at}$ if chemical potential $\mu$ equals zero. 

To build up the theory, we suppose an adiabatic switch of the hybridization at infinity.
Formally this means that $\Delta$ is multiplied by a slow profile function 
$\Delta^{\mathrm{prof}}(t)$  such as $\Delta^{\mathrm{prof}}(\pm \infty)=0$ and 
$\Delta^{\mathrm{prof}}(t)=1$ otherwise.
So, the evolution starts and finishes with a bare atom having a single electron. 
The complete evolution operator ${\cal S}(-\infty, \infty)$ preserves the spin orientation.
Therefore, it can be divided into the two parts, responsible 
for the evolution that starts and finishes with a certain spin orientation: 
\begin{equation}\label{SymmetryBreak}
 {\cal S}={\cal S}_{\up\up}+{\cal S}_{\down\down},
\end{equation}
where ${\cal S}_{\up\up}$ and ${\cal S}_{\down\down}$ describe the evolution 
$|\up\rangle_{-\infty}\rightarrow |\up\rangle_{+\infty}$ and 
$|\down\rangle_{-\infty}\rightarrow |\down\rangle_{+\infty}$.
Formally, ${\cal S}_{\sigma\sigma}$ can be defined as 
\begin{equation}
 {\cal S}_{\sigma\sigma}=\int_{|\sigma\rangle}^{|\sigma\rangle} e^{i S} \mathcal{D}[\bar c c],
\end{equation}
where $\int_{|\sigma\rangle}^{|\sigma\rangle}\mathcal{D}[\bar c c]$ implies integration over the
trajectories starting and finishing with a definite spin orientation.
Such a trick is necessary because of the degeneracy of the ground state of the atomic Hamiltonian.
Remind that for usual stationary diagrammatic technique, the ground state is unique,
and therefore always evolves to itself at infinite time. 

In practice, it is enough to consider only one part of the evolution operator, for example 
$\mathcal{S}_{\up\up}$. Clearly, the account of another part is equivalent to spin-averaging 
of the results obtained.

\section{Preliminary analysis}

A transformation to the dual variables \cite{dual_PRB2} requires a splitting of the action
into two parts. The first part might be nonlinear but should be exactly solvable, and the second
part should be Gaussian. The simplest way is to choose $S_{at}$ and hybridization as those parts.
Such a theory is described in the next two sections. It will be shown that it describes the
low-energy physics quite good. However, it will be shown also that a correct description at all
frequencies requires a more sophisticated approach, and a renormalization procedure will be
introduced.

Hubbard-Stratonovich decoupling of the hybridization and subsequent integration over $\bar c, c$
in the evolution operator results in the dual action 
\begin{equation}
 S[\bar f, f]=  \sum_{\sigma}\int\limits_{-\infty}^{+\infty} d\epsilon \left(\Delta^{-1}_{\epsilon} g_{\sigma\sigma,\epsilon}^{-2} - g_{\sigma\sigma,\epsilon}^{-1}\right) 
                \bar f_{\sigma,\epsilon} f_{\sigma,\epsilon} + V[\bar f,f],
\end{equation}
where Taylor coefficients of the nonlinear part $V[\bar f, f]$ are vertexes of the atomic problem. 

\begin{figure}
\onefigure[width=230pt]{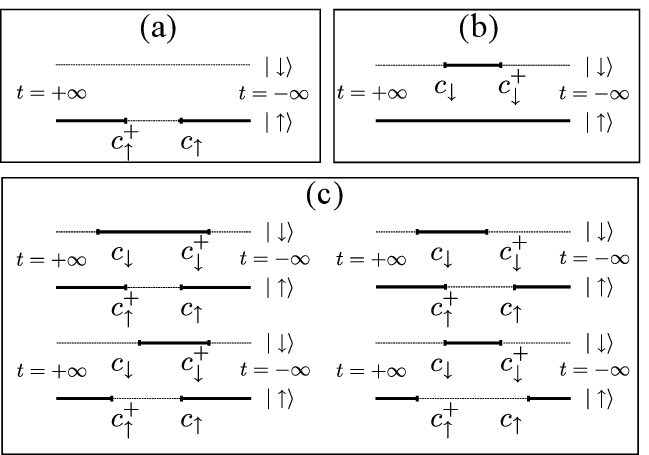}
\caption{The world-lines contributing to $g_{\sigma\sigma}$ (a,b)
	and $\gamma_{\up\down\up\down}^{(4)}$ (c).}
\label{worldlines}
\end{figure}

Momenta of the atomic problem can be calculated in the time domain, using world-line representation.
Parts (a) and (b) of Figure \ref{worldlines} show the nonvanishing world lines describing the
calculation of the Green's function. One can see, that since the evolution starts and finishes
with a single electron with the spin-up orientation, there is a definite ordering of the 
creation-annihilation operators: annihilation should precede creation for spin-up and 
vice versa for spin-down operators. 
At half-filling one obtains $g_{\up\up}(t) = i\theta(-t) e^{i(U/2)t}$ and 
$g_{\down\down}(t) = -i\theta(t)e^{-i(U/2)t}$ ($\theta$ is a Heaviside step function).
Fourier transform gives
\begin{equation}\label{Gatom}
 g_{\up\up}(\epsilon)=\frac{1}{\epsilon + U/2 - i 0}, ~~~ g_{\down\down}(\epsilon) = \frac{1}{\epsilon - U/2 + i 0}.
\end{equation}

These formulas are sufficient to construct the mean-field approximation, as 
higher momenta does not enter the theory in that case.
Very similarly to formulas (14-15) of \cite{dual_PRB2}, one obtains 
\begin{equation}\label{MF}
\begin{array}{c}
 {\cal G}_{\sigma\sigma}(\epsilon)=\left(g^{-1}_{\sigma\sigma}(\epsilon)-\Delta(\epsilon)\right)^{-1}\\
 {\cal G}^{dual}_{\sigma\sigma}(\epsilon)={\cal G}_{\sigma\sigma}(\epsilon)-g_{\sigma\sigma}(\epsilon)
\end{array}
\end{equation}
(through the paper, the mean-field results are denoted with the calligraphic letters).

Let us remind that an average over spin indices should be taken to obtain the final result,
so we get
\begin{equation}
 {\cal G}(\epsilon)=\frac{1/2}{\epsilon+U/2-\Delta(\epsilon)-i 0}+\frac{1/2}{\epsilon-U/2-\Delta(\epsilon)+i 0}
\end{equation}
(In our notation, quantities before and after spin averaging are marked with the same letters.
To escape confusion, we always supply non-averaged quantities with spin indices.)

We will be mostly interested in Kondo-like problem, so that atomic resonances $\pm U/2$ lie outside
the conduction band. One can see the  mean-field DOS of such a problem is built of slightly 
reshaped band DOS and two $\delta$-peaks at $\pm U/2$. The main fault of this approximation is
that no Kondo peak appears near Fermi level, while the higher-energy part of DOS is qualitatively
correct.
\section{Low-energy properties: general consideration}

A calulation of the corrections to the mean-field theory requires knowledge of higher momenta of
the atomic problem. The two-particle Green's function 
$g^{(2)}_{1234} = \langle c_1 c_2 c^\dag_3 c^\dag_4\rangle$ can be calculated similarly
to $g_{12}$ (indices here are combinations of energy and spin, for instance 1 stands for $\epsilon_1, \sigma_1$). 
The expression for $g^{(2)}$ contains 24 different terms, corresponding to various mutual order 
of the four time and spin arguments. However it turns out that only four of whose terms  contribute
to the non-Gaussian part $\Gamma_{1234} = g^{(2)}_{1234} - g_{13} g_{24} + g_{14} g_{23}$.
The world lines corresponding to these terms are depicted  in part (c) of Figure \ref{worldlines}.
After the Fourier transform (omiting the energy-conserving delta function), 
we obtain a simple formula for the fourth-order vertex 
$\gamma^{(4)}_{1234} = -i g^{-1}_{11} g^{-1}_{22} g^{-1}_{33} g^{-1}_{44} \Gamma_{1234}$.
Four-point vertex for all-the-same spin indices equals zero, and for the different indices 
\begin{equation}\label{gamma4}
  \gamma^{(4)}_{\up\down\up\down}(\epsilon_1,\epsilon_2;\epsilon_3,\epsilon_4) = 
    -U - \frac{U^2}{\epsilon_3-\epsilon_2 - 0 i}
\end{equation}
The first term of this expression is local in time, whereas the second one is proportional to
$\Theta(t_3-t_1) \delta(t_4-t_1) \delta(t_3-t_2)$ in time-domain.
The Heaviside function appears here because of the degeneracy of the ground state:
a time interval between the pairs $t_1 t_4$ and $t_2 t_3$ can be arbitrary large 
(see Figure \ref{worldlines}).

In this paper, we restrict ourselves with the simplest approximation beyond mean-field, that is the
first-order diagram correction to the dual self-energy:
\begin{equation}\label{Sigma1}
 \Sigma^{dual}_{\up\up}(\epsilon)=\frac{i}{2 \pi} \int \gamma^{(4)}_{\up\down\up\down} (\epsilon, \epsilon';\epsilon,\epsilon') {\cal G}^{dual}_ {\down\down}(\epsilon')  d\epsilon'
\end{equation}
and similarly for $\Sigma^{dual}_{\down\down}$.
It would be more accurate to use the integrand with the renormalized dual Green's function
$G^{dual}_ {\down\down}$, but we will stay with the expression with ${\cal G}^{dual}_ {\down\down}$
for simplicity.

To obtain an expression for the Green's function, it's practical to use the identity (18) of \cite{dual_PRB2} . We obtain
\begin{equation}\label{G1}
  G_{\sigma\sigma}(\epsilon) =
	\frac{1}{g^{-1}_{\sigma\sigma}(\epsilon)-\Delta(\epsilon)-
	\left(g_{\sigma\sigma}(\epsilon)+\left(\Sigma^{dual}_{\sigma\sigma}(\epsilon)\right)^{-1}\right)^{-1}}
\end{equation} 

An important property of these formulas is that $\Sigma^{dual}_{\sigma\sigma}$ contains the Kondo
logarithm. Indeed, substituting (\ref{gamma4}) into (\ref{Sigma1}) and taking into account that
$\Im {\cal G}^{dual}$ changes its sign at the Fermi level, one finds that $\Re \Sigma^{dual}_{\up\up}$ 
contains the logarithmic singuliarity:
$(2\pi)^{-1} \int\frac{U^2}{\epsilon - \epsilon' -i0} 
\Im {\cal G}^{dual}_{\down\down}(\epsilon') d\epsilon' \approx \pi^{-1} U^2 \log(-\Omega/\epsilon)
\Im {\cal G}^{dual}_{\down\down}(-0)$,
where the cut-off $\Omega$ is about the half-bandwidth. Futher substitution of Eq. (\ref{MF}) gives
an estimation $\Re \Sigma^{dual}_{\up\up} \approx 2\pi^{-1}\log(-\Omega/\epsilon)$ for the case
of large $U$. Clearly, such a logarithmic behaviour is also reflected in DOS.

The dual perturbation theory is valid while the corrections to the mean-field result are small enough.
According to formula (\ref{G1}), it means that the domain of validity is determined by the
inequality $g_{\up\up}\Sigma_{\up\up} \ll 1$. Note that at the Kondo energy \cite{hewson}
$\epsilon_K \propto \exp(-\frac{\pi U}{8})$ the left-hand side of this inequality equals one-half, so
our theory is formally valid only above the Kondo scale. But it is important to note that some
of low-energy physics is also reproduced. Namely, Friedel sum rule \cite{hewson} is satisfied:
as one can see from (\ref{G1}), the divergence of $\Sigma^{dual}_{\sigma\sigma}$ corresponds to
$G_{\sigma\sigma} = -\Delta^{-1}$.

A calculation for a concrete system should be presented to ensure that the theory behaves well. 
In the next section, we present the analytical formulas and corresponding graphs for the case
of half-filled system in a semicircular bath.
\section{Low-energy properties: analytical  results for a semicircular bath}

An explicit calculation of the first-order diagram in dual variables is
a simple although not trivial task. The bare dual Green's function ${\cal G}_{\up\up}^{dual}(\epsilon)$ has
an atomic pole at $-U/2$ (and at $U/2$ for ${\cal G}_{\down\down}^{dual}(\epsilon)$) as well as a branch cut
originating from the fact that $\Delta(\epsilon)$ represent a system with a continuous spectrum.

In our calculation we will use a semicircular $\Delta(\epsilon)$:
\begin{equation}
	\Delta(\epsilon) = \frac{1}{2}\left(\frac{2 t}{D}\right)^2
		(\epsilon - \sgn(\epsilon)\sqrt{\epsilon^2 - D^2})
\end{equation}

Here $D$ is the half-bandwidth, which for simplicity is put to one in all calculations below. 
Hybridization constant $t$ describes the coupling between the impurity atom and its nearest neighbor.

The diagram we are going to calculate is an integral over the real axis:
\begin{eqnarray}
\label{Sigma_uu}\Sigma_{\up\up}^{dual}(\epsilon) = -i\int \frac{d\epsilon'}{2\pi}\left(U + \frac{U^2}{\epsilon-\epsilon'-i0}\right) 
		{\cal G}_{\down\down}^{dual}(\epsilon') \\
\label{Sigma_dd}\Sigma_{\down\down}^{dual}(\epsilon) = -i\int \frac{d\epsilon'}{2\pi}\left(U + \frac{U^2}{\epsilon'-\epsilon-i0}\right) 
		{\cal G}_{\up\up}^{dual}(\epsilon')
\end{eqnarray}
 
Let's examine the pole structure of the part proportional to $U$ (constant part) in the expression 
for $\Sigma_{\up\up}^{dual}(\epsilon)$. Denote this part $\Sigma_{U}^{dual}$,
\begin{equation}
	\Sigma_{U}^{dual}=-i\frac{U}{2\pi}\int
		\frac{\Delta(\epsilon')\ d\epsilon'}{(\epsilon'-U/2 + i0)(\epsilon'-U/2+i0-\Delta(\epsilon'))}
\end{equation}

There are a pole at $\epsilon'=U/2-i0$ inherited from the atomic problem and at most two poles
which are solutions of the equation $\epsilon - U/2 = \Delta(\epsilon)$. For our special choice 
of $\Delta(\epsilon)$ this equation is reducible to the following system:
\begin{multline}
\left\{
\begin{array}{ll}
	\Re \Delta(\epsilon) &= \epsilon - U/2 \\
	\Im \Delta(\epsilon) &= 0
\end{array}
\right. \Leftrightarrow \\
\left\{
\begin{array}{ll}
	\epsilon^2(1-4 t^2) + \epsilon U(2 t^2-1) + ((2 t^2)^2 + (U/2)^2) = 0\\
	|\epsilon|>1
\end{array}
\right.
\end{multline}

\begin{figure}
\onefigure[width=160pt]{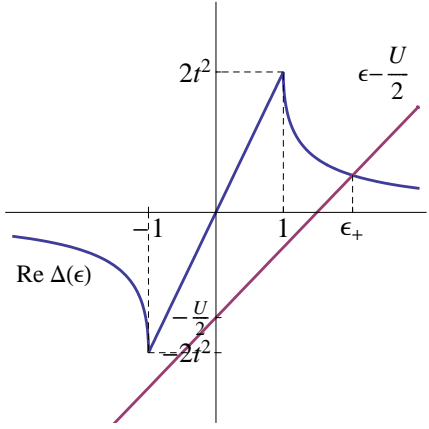}
\caption{Graphical solution of the equation $\Re \Delta(\epsilon) = \epsilon - U/2$.}
\label{pole_eqn}
\end{figure}

In the case of $t=1/2$ this quadratic equation turns to a linear one and its root
$\epsilon_0 = \frac{1}{2}\left(U+\frac{1}{U}\right)$ is always greater than or equal to 1.
When $t$ is not equal to 1/2, two real roots $\epsilon_+$ and $\epsilon_-$ are possible. 
This is a manifestation of an additional energy level splitting introduced by the 
''defect'' of the lattice. However in the limit $U \gg 2 t$ the only relevant root is 
still $\epsilon_+$. To understand this statement it's useful to represent
$\Re \Delta(\epsilon) = \epsilon - U/2$ equation in a graphical form (see Figure \ref{pole_eqn}). 

As one can see, the magnitude of $t$ determines the height of ``shoulders'' in this figure,
while $U/2$ is a vertical displacement of the diagonal line $\epsilon - U/2$. At $U$ large enough
in comparsion with $2 t^2$ (dimensionless) there is only one intersection of the curves on the positive half-axis,
i.e. $\epsilon_+$. A complete analisys of the equation yields the following results:
\begin{multline}
\begin{array}{l|l}
t,U & \textrm{roots} \\
\hline
t = 1/2 & \epsilon = \frac{1}{2}\left(U + \frac{1}{U}\right)\\
t\neq1/2, ~2 t^2 < 1, U/2 < 1 - 2 t^2 & \textrm{no real roots}\\
t\neq1/2, U/2 > |1 -2 t^2| & \epsilon=\epsilon_+\\
2 t^2 \ge 1, ~U/2 \le 2 t^2 - 1 & \epsilon=\epsilon_+,\epsilon_-
\end{array}\\
\epsilon_\pm = \frac{(U/2)(4 t^2 - 2) \pm 4 t^2\sqrt{4 t^2 + (U/2)^2 - 1}}{2(4 t^2 - 1)}
\end{multline}

\begin{figure}
\onefigure[width=160pt]{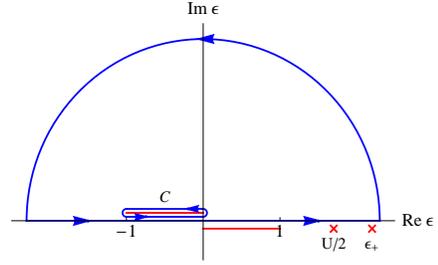}
\caption{A contour of integration for the constant part of $\Sigma_{\up\up}^{dual}(\epsilon)$.}
\label{contour}
\end{figure}

Both poles $U/2 - i0$ and $\epsilon_+ - i0$ reside in the fourth quadrant while $\Delta(\epsilon)$
has a branch cut on $[-1;1]$. To preserve casuality it has to perform an infinitesimal
shift up where $\epsilon<0$ and down for $\epsilon>0$. Such a disposition of the poles permits us to
choose an integration contour in the upper half-plane (the integrand vanishes at infinity as $\epsilon'^{-3}$)
and then deform it to a smaller contour $C$ as shown in Figure \ref{contour}.

Futher, the integrand of $\Sigma_{U}^{dual}$ can be split into two parts as follows

\begin{multline*}
	\Sigma_{U}^{dual} =\\
	-i\frac{U}{2\pi} \oint_C d\epsilon'
		\left(\frac{1}{\epsilon'-U/2+i0-\Delta(\epsilon')} - \frac{1}{\epsilon'-U/2+i0}\right)
\end{multline*}

One can observe that the second term of the integrand does not contribute to the result, since it 
contains no singuliarities in the contour.
To proceed with the first term, we substitute the explicit expressions for $\Re\Delta(\epsilon')$
and $\Im\Delta(\epsilon')$. We take into account that only $\Im\Delta(\epsilon')$  changes its sign
under a transition from one side of the branch cut to another. After routine transformations we obtain
\begin{multline}\label{final_integral}
\Sigma_{U}^{dual} =
	\frac{U}{2\pi}\frac{4 t^2}{1-4 t^2}\int_{-1}^0\frac{ \sqrt{1-\epsilon'^2} d\epsilon'}
		{(\epsilon'-\epsilon_+)(\epsilon'-\epsilon_-)}
\end{multline}

The integral can be simply done by a trigonometric substitution $\epsilon'=\sin\phi$.
So we obtain the final answer:
\begin{multline}
	\Sigma_{U}^{dual} =
	\frac{U}{2\pi}\frac{4 t^2}{1-4 t^2}\left(-\frac{\pi}{2} 
	+ \frac{L(\epsilon_+) - L(\epsilon_-)}{\epsilon_+ - \epsilon_-}\right)	
\end{multline}
\begin{equation}
	L(x) \equiv \sqrt{1-x^2}\log\left(\frac{\sqrt{1-x^2}+x-1}{\sqrt{1-x^2}-x+1}\right)
\end{equation}

There is no need to repeat all the calculations for the second part (proportional to $U^2$) of
equation (\ref{Sigma_uu}). An additional multiplier $(\epsilon-\epsilon'-i0)^{-1}$ produces another 
pole below the real axis, so it doesn't affect the integration contour in any way. This means that 
the last integral in (\ref{final_integral}) should be replaced with
\begin{multline*}
	\int_{-1}^0\frac{d\epsilon' \sqrt{1-\epsilon'^2}}
		{(\epsilon-\epsilon'-i0)(\epsilon'-\epsilon_+)(\epsilon'-\epsilon_-)} = \\
	\frac{1}{\epsilon_+-\epsilon_-}
		\int_{-1}^0\frac{d\epsilon' \sqrt{1-\epsilon'^2}}
		{\epsilon-\epsilon'-i0}\left(\frac{1}{\epsilon'-\epsilon_+} - \frac{1}{\epsilon'-\epsilon_-}\right)
\end{multline*}

In this way we have reduced the $U^2$-part to a known result. A full expression for 
$\Sigma_{\up\up}^{dual}(\epsilon)$ is
\begin{multline}\label{Sigma_uu_final}
	\Sigma_{\up\up}^{dual}(\epsilon) = \frac{U}{2\pi}\frac{4 t^2}{1-4 t^2}
		\left(-\frac{\pi}{2} + \frac{L(\epsilon_-) - L(\epsilon_+)}{\epsilon_- - \epsilon_+}
	\right. + \\\left.
	\frac{U}{\epsilon_+-\epsilon_-}
		\left(\frac{L(\epsilon-i0)-L(\epsilon_-)}{\epsilon-\epsilon_-} - 
		\frac{L(\epsilon-i0)-L(\epsilon_+)}{\epsilon-\epsilon_+}
	\right)\right) 
\end{multline}

(An infinitesimal imaginary part in $L(\epsilon-i0)$ assists to choose the right side of the branch
cut. Such an imaginary addition is not required for $\epsilon_\pm$, since they do not lie on the 
brunch cut for any positive $t$ and $U$).

An evaluation of $\Sigma_{\down\down}^{dual}(\epsilon)$ repeats the previous one with a number of exceptions:
\begin{itemize}
	\item Poles of the integrand are at $-U/2 + i0$ and $-\epsilon_+ + i0$.
	\item The integration contour goes in the opposite direction and surrounds $[0;1]$ segment.
	\item The pole of $(\epsilon'-\epsilon-i0)^{-1}$ lies above the real axis.
\end{itemize}
So we have a very similar answer:
\begin{equation}
	\Sigma_{\down\down}^{dual}(\epsilon) = -\Sigma_{\up\up}^{dual}(-\epsilon)
\end{equation}

By taking a limit in formula (\ref{Sigma_uu_final}) we achieve even simplier expression for the $t=1/2$ case:

\begin{multline}
    \Sigma_{\up\up}^{dual}(\epsilon)|_{t=1/2} = -\Sigma_{\down\down}^{dual}(-\epsilon)|_{t=1/2} = \\
        = \frac{\epsilon_0 - U}{2} - \frac{1}{\pi}\left(
         1 + L(\epsilon_0) -
        U\frac{L(\epsilon_0) - L(\epsilon - i0)}{\epsilon_0 - \epsilon}
        \right)\\
    \epsilon_0 = \frac{1}{2}(U + 1/U)
\end{multline}

The result for dual self energy should be substituted into formula (\ref{G1}). We have plotted thus
obtained DOS for several values of $U$ and obtained  smooth graphs with a clear Kondo-like peak at 
the Fermi level. 
So, it is  clearly demonstrated that the simplest first-order theory reproduces the low-energy physics 
of the Anderson impurity model correctly. However, we found that it suffers serious problems at higher
frequencies. This issue is discussed in the next section.

\section{Renormalization procedure}

The mentioned drawbacks of the formalism presented so far are related to the poles of 
$g_{\sigma\sigma}(\epsilon)$. First of all, as it follows from Eq. (\ref{G1}), the Green's
function $G_{\sigma\sigma}$ is pinned at its mean-field value $-1/\Delta$ at these points. The only
case when this pinning is absent is the vanishing of $\Sigma^{dual}_{\sigma\sigma}$ at the point of 
the pole, but it seems that no finite-order approximation for $g_{\sigma\sigma}(\epsilon)$ provides
such a condition. The pinning is rather unphysical, because poles of the atomic problem are 
by no means special points for the entire action (\ref{act}), and there is no any sum rule about
them. Further analysis shows that the theory fails also near the poles of the atomic problem:
it replaces the mean-field pole $\epsilon=U/2-\Delta(\epsilon)$ with the two poles shifted from
the real axis in different directions. Consequently, the theory is not conservative and not causal. 

The problems about the poles of $g_{\sigma\sigma}(\epsilon)$ are probably related to 
the pole structure of the dual Green's function. As it follows from the second line of (\ref{MF}),
it has two close poles, one placed exactly at $U/2$ and another slightly shifted from this point.
The residues of these poles have opposite signs. We suppose that the description using self-energy
is not valid in this situation. Anyhow, the above paragraph contains a constructive idea on how to 
improve the theory. Since the  pinning is absent only if dual self-energy vanishes at the pole,
\begin{equation} \label{require}
G_{\sigma\sigma}(\epsilon_\mathrm{pole})=0; \quad g^{-1}_{\sigma\sigma}(\epsilon_\mathrm{pole})=0,
\end{equation} 
its reasonable to {\it require} the fulfillment of this condition. An additional condition means
that the theory must have an additional ajustable parameter. To introduce it, we modify the
splitting of the action into Gaussian and Hamiltonian parts. We rewrite (\ref{act}) as follows:
\begin{equation}\label{act2}
\begin{array}{r}
S = {S'}_{at} -\iint\limits_{-\infty}^{+\infty} dt dt' \bar c_{\sigma t} \left(\Delta(t-t') +i \lambda \delta'(t-t') \right) c_{\sigma t'}\\   \\
S'_{at} = \int\limits_{-\infty}^{+\infty} dt \left( 
        i (1-\lambda) \bar c_{\sigma t} \frac{\partial}{\partial t} c_{ \sigma t} 
           -  U \tilde{n}_{\up t} \tilde{n}_{\down t}   +\mu \tilde{n}_{\sigma t}
            \right)
\end{array} 
\end{equation}
The parameter $\lambda$ is to be adjusted to maintain the condition (\ref{require}).

The calculations are very similar to the above considered case $\lambda=0$. 
The atomic problem is the same, up to scaling transformations. We obtain
\begin{equation}
 \begin{array}{l}
 g_{\sigma\sigma}(\epsilon) = \left((1-\lambda) \epsilon\pm U'/2\mp i0\right)^{-1};
\quad U' \equiv (1-\lambda)^{-1}U \\
\gamma^{(4)}_{\up\down\up\down}(\epsilon_1,\epsilon_2;\epsilon_3,\epsilon_4) =
- U  -  \frac{U'^2}{\epsilon_3-\epsilon_2 - 0 i}\\
\end{array}
\end{equation} 
The mean-field consideration gives
\begin{equation}\label{MFrenorm}
    {\cal G}_{\sigma\sigma}(\epsilon)=\left(\epsilon\pm U' -\Delta(\epsilon)\mp i 0\right)^{-1},
 \end{equation}
so that the renormalization just results in a shift of the atomic resonances. 
Practical calculation shows that (\ref{require}) is fulfilled with a small positive $\lambda$,
and that $\lambda$ grows as $U$ decreases.
 
A calculation of the first-order correction is also similar. 
The only important difference is that ${\cal G}^{dual}_{\sigma\sigma}(\epsilon)$ obeys high 
frequency asymptotics $\frac{\lambda}{1-\lambda}\epsilon^{-1}$ for a finite $\lambda$,
and therefore the contour integration contains an additional contribution from infinity.
The final formula (\ref{Sigma_uu_final}) becomes:
\begin{multline}
	\Sigma_{\up\up}^{dual}(\epsilon;\lambda) = -\Sigma_{\down\down}^{dual}(-\epsilon;\lambda) =\\
	=\frac{U}{2\pi}\frac{4 t^2}{1-4 t^2}
		\left(-\frac{\pi}{2} + \frac{L(\epsilon'_-) - L(\epsilon'_+)}{\epsilon'_- - \epsilon'_+}\right) + \\
	\frac{U'^2}{2\pi}\frac{4 t^2}{1-4 t^2}\frac{1}{\epsilon '_+-\epsilon'_-}
		\left(\frac{L(\epsilon-i0)-L(\epsilon'_-)}{\epsilon-\epsilon'_-} - \right. \\ \left. 
		\frac{L(\epsilon-i0)-L(\epsilon'_+)}{\epsilon-\epsilon'_+}
	\right) + \lambda\frac{U'}{2} \\
	\epsilon'_\pm = \frac{(U'/2)(4 t^2 - 2) \pm 4 t^2\sqrt{4 t^2 + (U'/2)^2 - 1}}{2(4 t^2 - 1)}
\end{multline} 

\begin{figure}
\onefigure[width=0.9\columnwidth]{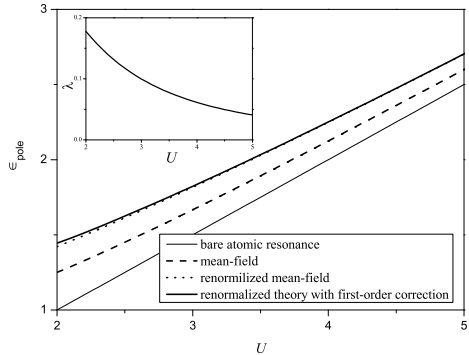}
\caption{Position of the atomic resonance of SIAM with $t=D/2$ in different approximations. 
The resonance bare atom (that is, $\epsilon_{pole}=\frac{U}{2}$ ) is shown for comparison.
Note that the renormalized theory gives a twice-larger shift of the resonance position than the
mean-field approach does. For the renormalized theory, $\Sigma$ vanishes near the resonance
point, therefore taking the dual correction into account almost does not affect the resonance.
Inset shows the value of renormalization parameter $\lambda$.}
\label{pole}
\end{figure}

\begin{figure}
\onefigure[width=0.9\columnwidth]{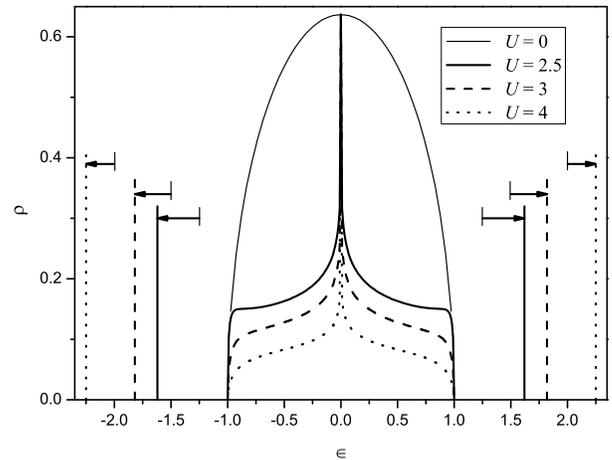}
\caption{Local density of states $\rho(\epsilon) = -\frac{\sgn(\epsilon)}{\pi}\Im G(\epsilon)$ 
of SIAM with $t=D/2$, calculated from the renormalized theory. Atomic resonances are shown
with vertical lines; height of the line is determined by the spectral weight of the resonance.
Arrows indicate shift from the resonance of the bare atom.}
\label{DOS}
\end{figure}

We have performed calculations with this dual self-energy substituted in Eq. (\ref{G1}).
The value of $\lambda$ has been ajusted numerically to fulfill the condition (\ref{require}).
The resulting graphs for the band DOS are shown in Figure \ref{DOS}.
Besides band DOS, there are also separated poles, corresponding to the atomic resonances.
Their positions are shown in Figure \ref{pole}. For comparison, we also plot the mean-field values
with and without renormalization (poles of the expressions (\ref{MFrenorm}) and (\ref{MF}),
respectively). It can be verified, that the renormalized theory is causal (that is, poles
are infinitesimally shifted from the real axis in proper quadrants) and conservative (that is,
Luttinger theorem is fulfilled).

\section{Conclusion}
At the end of the paper we can compare our scheme to the approach by Logan {\it et al}
\cite{logan}, since the local moment approach developed there is formally similar to
the present theory in a number of points. A similar form of the bare Green's functions
is used to construct a diagramatic expansion. But the Green's functions of the local moment
approach are to be determined self-consistently, so that it includes poles of a partially
``dressed'' impurity atom. In the same fashion summation of ladder diagrams is required to
obtain a renormalized energy-dependent vertex having a neccesary peculiarity near zero
frequency. So, the approach by D. Logan requires an essential numerics. In contrast, we
presented an almost analytical scheme. Therefore, the result can be improved in a regular way
by the account of higher-order diagrams. Another important peculiarity of our approach is
that it explicitly describes the renormalization of atomic levels due to hybridization with
conduction electrons. A crucial checkpoint for the further development of the method
is its applicability to multi-orbital systems. As far as we know, D. Logan's approach met
serious difficulties beyond single-band systems, although there is a promising result of the
variational scheme \cite{byczuk}. For our scheme multi-orbital generalization looks straightforward.

The work was supported by RFFI-DFG grant 08-02-91953 --- 436-RUS-113-938-0.

\bibliographystyle{eplbib.bst}
\bibliography{article.bib}

\end{document}